# Comment on "Using Dipole Interaction to Achieve Nonvolatile Voltage Control of Magnetism in Multiferroic Heterostructures" [Adv. Mater., 2021, 2105902]

Supriyo Bandyopadhyay
Department of Electrical and Computer Engineering
Virginia Commonwealth University, Richmond, VA 23284, USA
Email: sbandy@vcu.edu

In a recent article[1], A. Chen et al. claimed that they have switched (rotated by ~$90^0$) the magnetization of the soft layer of a magnetic tunnel junction (MTJ) in a *non-volatile* way with *volatile* strain by exploiting *dipole interaction* with the hard layer. I show that dipole interaction cannot cause this effect and the authors' explanation cannot be correct. An alternate explanation is offered.

The authors applied a voltage-generated strain to the soft layer of an MTJ (using a piezoelectric substrate) to rotate its magnetization away from the easy axis to the hard axis. When they withdrew the voltage, the magnetization remained pointing along the hard axis and did not relax back to the easy axis until a voltage of opposite polarity was applied (hence the non-volatility). The authors posit that this happened because the dipole interaction between the soft and the hard layers of the MTJ prevented the magnetization of the soft layer from relaxing back to its easy axis upon withdrawal of the voltage.

Dipole interaction cannot cause this to happen. To see why, consider the rectangular soft layer (CoFeB) of the MTJ shown in Fig. 1, where the hard axis is along the *x*-direction and the easy axis is along the *y*-direction (the same convention as that used by the authors). The discussion here will be restricted to the energetics in the plane of the soft layer since the magnetic layers are thin and have in-plane anisotropy. Assume that the magnetization subtends an angle θ with the easy axis as shown in Fig. 1. The potential energy in the plane of the soft layer as a function of magnetization orientation, under four relevant conditions, are shown schematically in Figs. 1(a) – 1(d), assuming there is no magneto-crystalline anisotropy. The magnetic layers are not single crystals and hence there is no magneto-crystalline anisotropy to consider (which is why Chen et al. also did not mention it). This is a schematic depiction which is not exact but that does not detract from the ensuing discussion. Note that *only under compressive strain*, the energy minimum is at θ = $90^0$ and only then the magnetization can point along the hard axis. If there is no compressive strain, dipole interaction *cannot* keep the soft layer's magnetization pointing along the hard axis because the energy minimum is then not at θ = $90^0$. Hence, dipole interaction cannot be responsible for the observed effect. In fact, without remanent compressive strain, the hard axis (θ = $90^0$) becomes the energy *maximum* (with or without dipole interaction) and if the magnetization remains pointing along the hard axis, then it will have to remain stable at the energy maximum (and not decay to the energy minimum) which violates the second law of thermodynamics. Thus, dipole interaction cannot explain the observed effect in ref. [1], but remanent strain can.

There are other indicators that support the notion of remanent strain being the cause rather than dipole interaction. According to the authors, +150 V generates uniaxial compressive strain and -150 V generates uniaxial tensile strain along the easy axis in the soft layer. The actual strain is very likely biaxial[2] (compressive along the easy axis and tensile along the hard axis for positive voltage and the opposite for negative voltage) but that is not important for this discussion since compressive (tensile) strain along the easy axis has the same effect on magnetization as tensile (compressive) strain along the hard axis. It is then very understandable that +150 V will make the magnetization point along the hard axis (*x*-axis) and -150 V will bring it back to the easy axis (*y*-axis) [as shown in figure 6 of ref. 1] because the energy minimum moves to $\theta = 90^0$ under compressive strain [+150 V] and to $\theta = 180^0$ under tensile strain [-150 V], as shown in Figs. 1(c) and 1(d). However, it is *not understandable* why the magnetization will remain pointing along the hard axis ($\theta = 90^0$) after the voltage is reduced to +0 V if indeed the strain vanishes at +0 V (no remanent compressive strain) because then the energy minimum will no longer be at $\theta = 90^0$, but move to $\theta = 180^0$ (in the presence of dipole interaction) as shown in Fig. 1(b). The magnetization simply cannot remain pointing along the hard axis at +0 V unless there is some remanent compressive strain at +0 V, because otherwise we have to accept the possibility that the system has become unconditionally stable at the energy maximum (in the presence of thermal noise) which violates the second law of thermodynamics. In this discussion, we have ignored the effect of any spurious states due to defects, etc. Those are random and uncontrollable in any case.

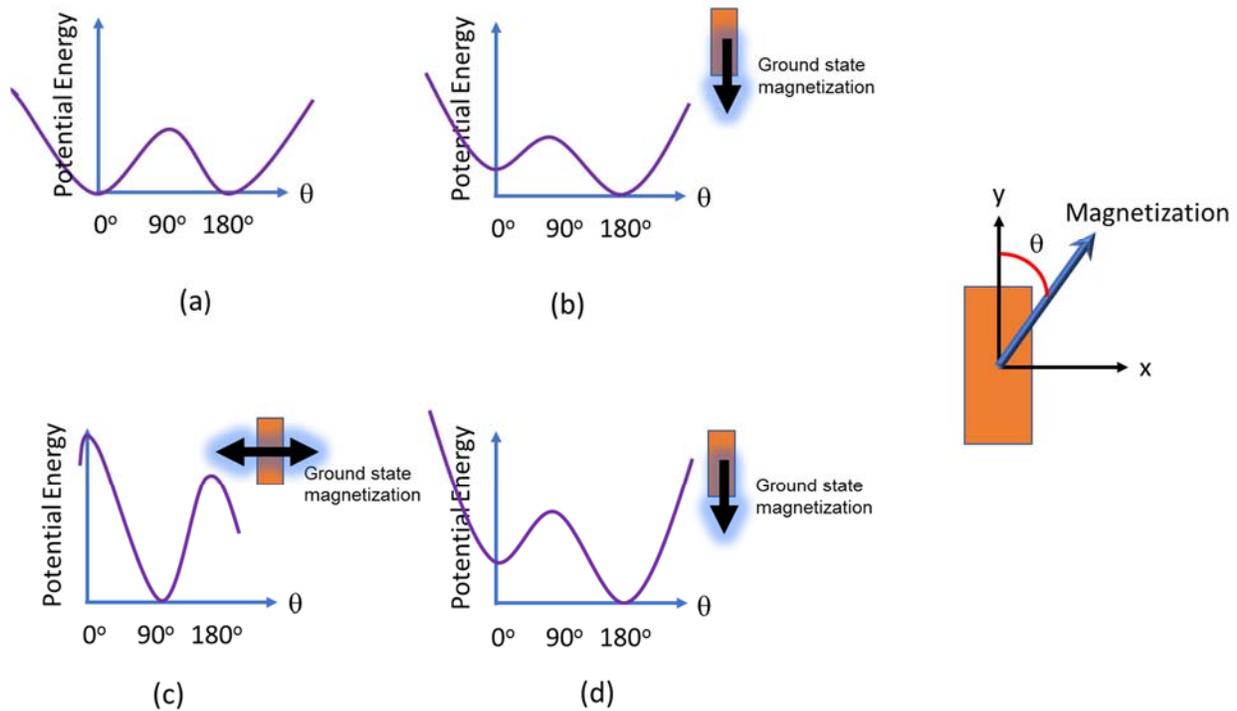

Fig. 1: Potential energy landscape for in-plane magnetization of the CoFeB soft layer in the absence of any magneto-crystalline anisotropy. (a) Only shape anisotropy is present, (b) shape anisotropy and dipole interaction with the hard Co layer are present [we assume that the hard layer's magnetization is along the +y-axis or $\theta = 0^0$], (c) shape anisotropy, dipole interaction with the hard Co layer, and large uniaxial compressive strain along the easy axis are present, and (d) shape anisotropy, dipole interaction with the hard Co layer, and uniaxial tensile strain along the easy axis are present. The shapes of the potential profiles here are approximate, but the exact shapes have no bearing on the discussion here.

It is therefore remanent compressive strain in the CoFeB soft layer at +0 V, and not dipole interaction with the Co layer, that keeps the magnetization pointing along the *x*-direction (hard axis) when the voltage is reduced to +0V. The remanent strain accrues from the residual strain in the piezoelectric which does not vanish when the voltage is reduced to zero. This has been reported by many authors[3-8]. Suffice it to say then that the non-volatility observed in ref. [1] has *nothing to do with dipole interaction* but is due to the soft layer *retaining strain in the absence of voltage*, meaning that it is the strain that is "non-volatile" and dipole interaction does not play a role in the non-volatility at all. A negative voltage generates strain of the opposite sign (tensile) and reverses the remanent strain in the soft layer which is why a negative voltage is needed for the reset operation.

The problem with this kind of non-volatility is that if the remanent strain in the soft layer relaxes spontaneously owing to temperature variation, etc., then the entire non-volatility would go away, making this type of non-volatility somewhat unstable. In any case, the claim in the abstract of ref. [1] that "nonvolatile voltage control … originates from the nonvolatile magnetization rotation of an <u>interacting</u> CoFeB magnet driven by <u>volatile</u> voltage-generated strain" is not tenable *because neither is the voltage-generated strain volatile, nor does the (dipole) interaction play a role in the non-volatility.*

The authors claim while discussing their figure 2 that since the *M-H* loops at +0 V and -0 V are almost identical for an *isolated* CoFeB nanomagnet, there cannot be any residual (non-volatile) strain in the isolated nanomagnet. First, the +0 V and -0 V M-H loops being nearly identical does not prove that there is no residual strain in an isolated nanomagnet. CoFeB is magnetostrictive and hence the external magnetic field used to generate the *M-H* loops can cause it to expand/contract during measurement and thus relax any residual strain while generating the *M-H* loops. Therefore, this measurement sheds no light on whether there is residual strain in the isolated nanomagnet. Second, even if there is no residual strain in the isolated nanomagnet, that does not mean there is none in the MTJ since the CoFeB layer has two different interfaces in the two cases. In fact, *because there may be residual strain in the soft layer of the MTJ* – compressive in the case of +0 V and tensile in the case of -0V – the MTJ resistance is different at +0 V and -0 V in the absence of any external magnetic field, which what the authors observed in their figure 3(e). The same causation explains their figure 3(c); the +0 V and -0 V curves are different because the remanent strain is compressive in one case and tensile in the other. Their figure 5 reports micromagnetic simulations based on the *assumption* of no residual strain and hence sheds no light on this question. It is entirely possible that had the authors included residual strain in their simulations, they would have observed the same non-volatility. In fact, if they had included residual strain and *excluded* dipole interaction, they would very likely have observed the same non-volatility again. Since these tests were not done, the simulations offer no support for the authors' conjecture that dipole interaction is the cause of the effect.

There are other minor concerns. The authors claim "with a decreasing size and distance between the magnets, dipole interaction between two adjacent magnets cannot be neglected". While decreasing separation indeed increases dipole interaction, decreasing size does just the opposite since the dipole interaction energy is proportional to the square of the nanomagnet volume (larger nanomagnets have stronger dipole interaction).

Second, the magnetization rotation demonstrated is through $90^0$ and not full $180^0$. This is usually not preferred since it reduces the ratio of the high-to-low resistance of the MTJ considerably. In ref. [9-11] which did leverage dipole interaction to flip magnetization, but needed a magnetic field for reset instead of the electric field, the magnetization was flipped by full $180^0$ and that leads to a much higher resistance ratio which is conducive to device applications.

Finally, there have been proposals and demonstrations of switching the magnetization of a nanomagnet in a *non-volatile* way using *volatile* strain, both via $90^0$ rotation[12] and via $180^0$ rotation[13]. These techniques do not need dipole interaction and can switch isolated nanomagnets. Dipole interaction is also usually not preferred in MTJs (which is why the hard layer is fashioned out of a synthetic antiferromagnet) because this interaction makes it hard to switch the MTJ from the antiparallel to the parallel state.